\documentclass[twocolumn]{aastex62}
\usepackage{comment}
\usepackage{amsmath}	
\usepackage{amssymb}
\usepackage{footnote}
\usepackage{appendix}
\usepackage{upgreek}
\usepackage{tablefootnote}
\usepackage[graphicx]{realboxes}
\usepackage{threeparttable}
\usepackage[caption=false]{subfig}

\graphicspath{{./}{figures/}}

\newcounter{daggerfootnote}
\newcommand*{\daggerfootnote}[1]{%
    \setcounter{daggerfootnote}{\value{footnote}}%
    \renewcommand*{\thefootnote}{\fnsymbol{footnote}}%
    \footnote[2]{#1}%
    \setcounter{footnote}{\value{daggerfootnote}}%
    \renewcommand*{\thefootnote}{\arabic{footnote}}%
    }
\published{in ApJL, September 24, 2020}

\shorttitle{Commensal ASKAP FRB}
\shortauthors{Bhandari et al.}

\begin{document}

\title{Limits on precursor and afterglow radio emission from a fast radio burst in a star-forming galaxy}

\correspondingauthor{Shivani Bhandari}
\email{shivani.bhandari@csiro.au}
\author[0000-0003-3460-506X]{Shivani Bhandari}
\affil{Australia Telescope National Facility, CSIRO Astronomy and Space Science, PO Box 76, Epping, NSW 1710, Australia }

\author[0000-0003-2149-0363]{Keith W. Bannister}
\affil{Australia Telescope National Facility, CSIRO Astronomy and Space Science, PO Box 76, Epping, NSW 1710, Australia }

\author{Emil Lenc}
\affil{Australia Telescope National Facility, CSIRO Astronomy and Space Science, PO Box 76, Epping, NSW 1710, Australia }

\author{Hyerin Cho}
\affiliation{School of Physics and Chemistry, Gwangju Institute of Science and Technology, Gwangju, 61005, Korea}

\author{Ron Ekers}
\affil{Australia Telescope National Facility, CSIRO Astronomy and Space Science, PO Box 76, Epping, NSW 1710, Australia }
\affiliation{International Centre for Radio Astronomy Research, Curtin University, Bentley WA 6102, Australia }
\author[0000-0002-8101-3027]{Cherie K.Day}
\affiliation{Centre for Astrophysics and Supercomputing, Swinburne University of Technology, John St, Hawthorn, VIC 3122, Australia}
\affiliation{Australia Telescope National Facility, CSIRO Astronomy and Space Science, PO Box 76, Epping, NSW 1710, Australia}

\author{Adam T.Deller}
\affiliation{Centre for Astrophysics and Supercomputing, Swinburne University of Technology, John St, Hawthorn, VIC 3122, Australia}

\author{Chris Flynn}
\affiliation{Centre for Astrophysics and Supercomputing, Swinburne University of Technology, John St, Hawthorn, VIC 3122, Australia}

\author[0000-0002-6437-6176]{Clancy W. James}
\affiliation{International Centre for Radio Astronomy Research, Curtin University, Bentley WA 6102, Australia }

\author[0000-0001-6763-8234]{Jean-Pierre Macquart$^\dagger$}
\affiliation{International Centre for Radio Astronomy Research, Curtin University, Bentley WA 6102, Australia }
\daggerfootnote{Deceased}

\author{Elizabeth K. Mahony}
\affil{Australia Telescope National Facility, CSIRO Astronomy and Space Science, PO Box 76, Epping, NSW 1710, Australia }

\author[0000-0003-1483-0147]{Lachlan Marnoch}
\affiliation{Department of Physics and Astronomy, Macquarie University, NSW 2109, Australia}
\affiliation{Macquarie University Research Centre for Astronomy, Astrophysics \& Astrophotonics, Sydney, NSW 2109, Australia}
\affiliation{Australia Telescope National Facility, CSIRO Astronomy and Space Science, PO Box 76, Epping, NSW 1710, Australia }

\author{Vanessa A. Moss}
\affil{Australia Telescope National Facility, CSIRO Astronomy and Space Science, PO Box 76, Epping, NSW 1710, Australia }
\affiliation{Sydney Institute for Astronomy, School of Physics, University of Sydney, NSW 2006, Australia}

\author{Chris Phillips}
\affil{Australia Telescope National Facility, CSIRO Astronomy and Space Science, PO Box 76, Epping, NSW 1710, Australia }

\author{J. Xavier Prochaska}
\affil{University of California, Santa Cruz, 1156 High St., Santa Cruz, CA 95064, USA}
\affiliation{
Kavli Institute for the Physics and Mathematics of the Universe (Kavli IPMU),
5-1-5 Kashiwanoha, Kashiwa, 277-8583, Japan}

\author{Hao Qiu}
\affiliation{Sydney Institute for Astronomy, School of Physics, University of Sydney, NSW 2006, Australia}
\affil{Australia Telescope National Facility, CSIRO Astronomy and Space Science, PO Box 76, Epping, NSW 1710, Australia }

\author[0000-0003-4501-8100]{Stuart D. Ryder}
\affiliation{Department of Physics and Astronomy, Macquarie University, NSW 2109, Australia}
\affiliation{Macquarie University Research Centre for Astronomy, Astrophysics \& Astrophotonics, Sydney, NSW 2109, Australia}

\author[0000-0002-7285-6348]{Ryan M. Shannon}
\affiliation{Centre for Astrophysics and Supercomputing, Swinburne University of Technology, John St, Hawthorn, VIC 3122, Australia}

\author[0000-0002-1883-4252]{Nicolas Tejos}
\affiliation{Instituto de F\'isica, Pontificia Universidad Cat\'olica de Valpara\'iso,
Casilla 4059, Valpara\'iso, Chile}

\author{O. Ivy Wong}
\affiliation{CSIRO Astronomy \& Space Science, PO Box 1130, Bentley, WA 6102, Australia}

\begin{abstract}

We present a new fast radio burst (FRB) at 920 MHz discovered during commensal observations conducted with the Australian Square Kilometre Array Pathfinder (ASKAP) as part of the Commensal Real-time ASKAP Fast Transients (CRAFT) survey. FRB~191001 was detected at a dispersion measure (DM) of 506.92(4) pc~cm$^{-3}$ and its measured fluence of 143(15) Jy~ms is the highest of the bursts localized to host galaxies by ASKAP to date. The subarcsecond localization of the FRB provided by ASKAP reveals that the burst originated in the outskirts of a highly star-forming spiral in a galaxy pair at redshift $z=0.2340(1)$. Radio observations show no evidence for a compact persistent radio source associated with the FRB~191001 above a flux density of $15\upmu$Jy. However, we detect diffuse synchrotron radio emission from the disk of the host galaxy that we ascribe to ongoing star formation. FRB~191001 was also detected as an image-plane transient in a single 10 s snapshot with a flux density of 19.3~mJy in the low-time-resolution visibilities obtained simultaneously with CRAFT data. The commensal observation facilitated a search for repeating and slowly varying radio emissions 8 hr before and 1 hr after the burst. We found no variable radio emission on timescales ranging from 1~ms to 1.4~hr. We report our upper limits and briefly review FRB progenitor theories in the literature that predict radio afterglows. Our data are still only weakly constraining of any afterglows at the redshift of the FRB. Future commensal observations of more nearby and bright FRBs will potentially provide stronger constraints.

\end{abstract}


\section{Introduction} \label{intro}

Fast radio bursts (FRBs) are energetic bursts of radio emission that last for tens of microseconds to tens of milliseconds \citep{Lorimer} and originate at cosmological distances \citep{VLAlocalisation, Bannister+19, Ravi+19}.
More than 20 FRB sources have been observed to emit repeat pulses \citep{spitler2016repeating, andersen2019chime, kumar2019faint}, allowing some of them to be localized to host galaxies via targeted follow-up with radio interferometers \citep{Tendulkar2016, Marcote2020}.

The bulk of the $\sim$100-strong population of published FRBs, however, are single-burst events. The transition from finding such events with single dishes to with interferometric arrays capable of imaging the received FRB emission has resulted in (sub-)arcsecond localization in recent detections  \citep{Bannister+19,Ravi+19, X+19, JP+20, Law2020}, revealing their host galaxies and in some cases even to sites within the hosts \citep{Bhandari_2020}.

Analyses of the host environments of localized repeating and non-repeating FRBs and high brightness temperatures of the bursts tend to favor models involving compact objects such as white dwarfs (WD), neutron stars (NS) and black holes (BH) \citep{Liu2018,Murase_2018,WangLai2020,Wang2020}. Some of these models predict radio afterglows accompanying an FRB with timescales of days to years and apparent luminosities as high as sub-mJy levels in favorable circumstances.
 
\begin{table}
 \caption{Measured and derived properties of FRB~191001 and its host galaxy. }
    \label{tab:properties}
    \begin{tabular}{ll}
        \hline
         Properties & \\
         \hline
         \textbf{FRB} \\
    
         Arrival time (UT) \footnote{The statistical uncertainty on the burst arrival time assumes a model for the burst morphology.} at 919.5~MHz & 16:55:35.97081 \\
         Incoherent S/N & 62\\
         Coherent S/N  & 192 \\
         Primary detection beam & 24 \\
         Detection DM (pc~cm$^{-3}$) & 506.92(4) \\
         Structure maximised DM\footnote{ \url{https://github.com/danielemichilli/DM_phase/blob/master/DM_phase.py}} (pc~cm$^{-3}$) & 507.90(7)\\
         DM$_{\rm ISM}~\rm NE2001 $ (pc~cm$^{-3}$) & 44 \\
         DM$_{\rm ISM}~\rm YMW16 $ (pc~cm$^{-3}$) & 31 \\
         DM$_{\rm cosmic}$ (pc~cm$^{-3}$) & $373_{(\rm NE2001)}$  \\
         RA (J2000) & 21:33:24.37(2) \\ DEC(J2000) & $-$54:44:51.86(13)\\
         Fluence\footnote{Derived from 336~MHz bandwidth CRAFT intensity data. } (Jy\,ms) & 143(15)\\
         Modelled pulse width\footnote{The primary pulse component width.} (ms)& 0.22(3)\\
         Scattering time at 824 MHz (ms) &  $3.3(2)$ \\
         Scintillation bandwidth (kHz) & $\sim600$ \\
         Pulse rise time ($\upmu$s) & $\sim 640$\\
         Rotation measure (RM) (rad~m$^{-2}$) & $55.5(9)$ \\
         Total polarisation fraction (P/I) & 58(1)\% \\
         Linear polarisation fraction (L/I) & 57(1)\% \\
         Circular polarisation fraction (V/I) & $-$5(1)\% \\
         Spectral energy density \tnote{a} (erg~Hz$^{-1}$) & $2 \times 10^{32}$\\
         Persistent source,  & 2 $\times 10^{21}$ \\
         radio luminosity at 5.5~GHz (W Hz$^{-1}$) & \\
         \\
         \textbf{Host galaxy}\footnote{See \citet{heintz2020} for details about optical properties.} \\
         Redshift & 0.2340(1)\\
        Stellar mass ($M_\odot$) & $5(2)  \times 10^{10}$ \\
        E(B-V) & 0.27(16)  \\
        \hline
      \end{tabular}
    
\end{table}   

\citet{Margalit2019} predicted late-time (months to years) radio emission accompanying FRBs from magnetars born in binary neutron star mergers (BNS) and accretion-induced collapse (AIC) of a white dwarf (WD) through the interaction of the ejecta with the surrounding local interstellar medium. It is also proposed that the magnetospheric instability of an isolated or a binary rotating black hole may result in FRBs and their afterglows \citep{Liu_2016}.
If FRBs are related to gamma-ray bursts (GRBs) \citep{Zhang2014}, the radio afterglow may last from hours to years as reverse and forward shocks interact with ejecta and the interstellar medium \citep{Frail2003}. \citet{Yi2014} also predicted the optical, radio and high energy afterglow light curves for forward and reverse shock emission resulting from FRBs. They concluded that FRB afterglows are too faint to be detected by current detectors.

While theoretical predictions for radio afterglows are plentiful, observational evidence is scanty. An initially promising candidate was identified from radio follow-up conducted within $\sim$~2hr of the Parkes FRB~150418, reporting fading radio emission in the field \citep{keane2016host}. Subsequent observations showed that scintillation of an unassociated active galactic nucleus (AGN) was a more plausible explanation \citep{WB,Johnston2017}.
In 2018, multi-wavelength follow-up of three real-time FRBs did not detect any afterglows in radio or other wavelengths \citep{Bhandari2018}.
Potential explanations include FRBs having fainter radio afterglows (less than $50\upmu$Jy at $3\sigma$), these afterglows evolving on timescales faster than what had been surveyed (less than a day), or FRBs not producing afterglows. Deeper follow-up observations of a larger sample of FRBs are needed to address the first possibility, but the second is best addressed through observations where radio imaging is being performed during the same observation in which the FRB is discovered.

Since mid-2019, FRB searches with the Commensal Real-time ASKAP Fast Transients Survey \citep[CRAFT]{mbb+10} have started to operate simultaneously with other survey science projects undertaken with ASKAP. In contrast to previously reported CRAFT detections \citep{Bannister+19,X+19,JP+20}, during commensal operations, the regular ASKAP correlator is still running and producing low-time-resolution visibility products needed for regular ASKAP imaging. The simultaneous operation of the ASKAP hardware correlator and ASKAP-CRAFT system is powerful in probing long-timescale radio emission before and after the FRB itself. FRB~191001 was the first event to be detected during such an observation, facilitating a deep search for slowly varying ($\geq10$-s) radio emission within hours before and after the FRB.
 
In this paper we report the detection and localization of FRB~191001 to its host galaxy along with a search for radio emission pre-- and post--FRB. The detection and properties of the FRB and its host galaxy are presented in Section 2, and the results of our search for radio emission from the host galaxy at the location of the FRB in Section 3. In Section 4, we discuss the implications of our findings, and we conclude and provide a summary in Section 5.  

\begin{figure*}
\begin{tabular}{cc}
\includegraphics[scale=0.52]{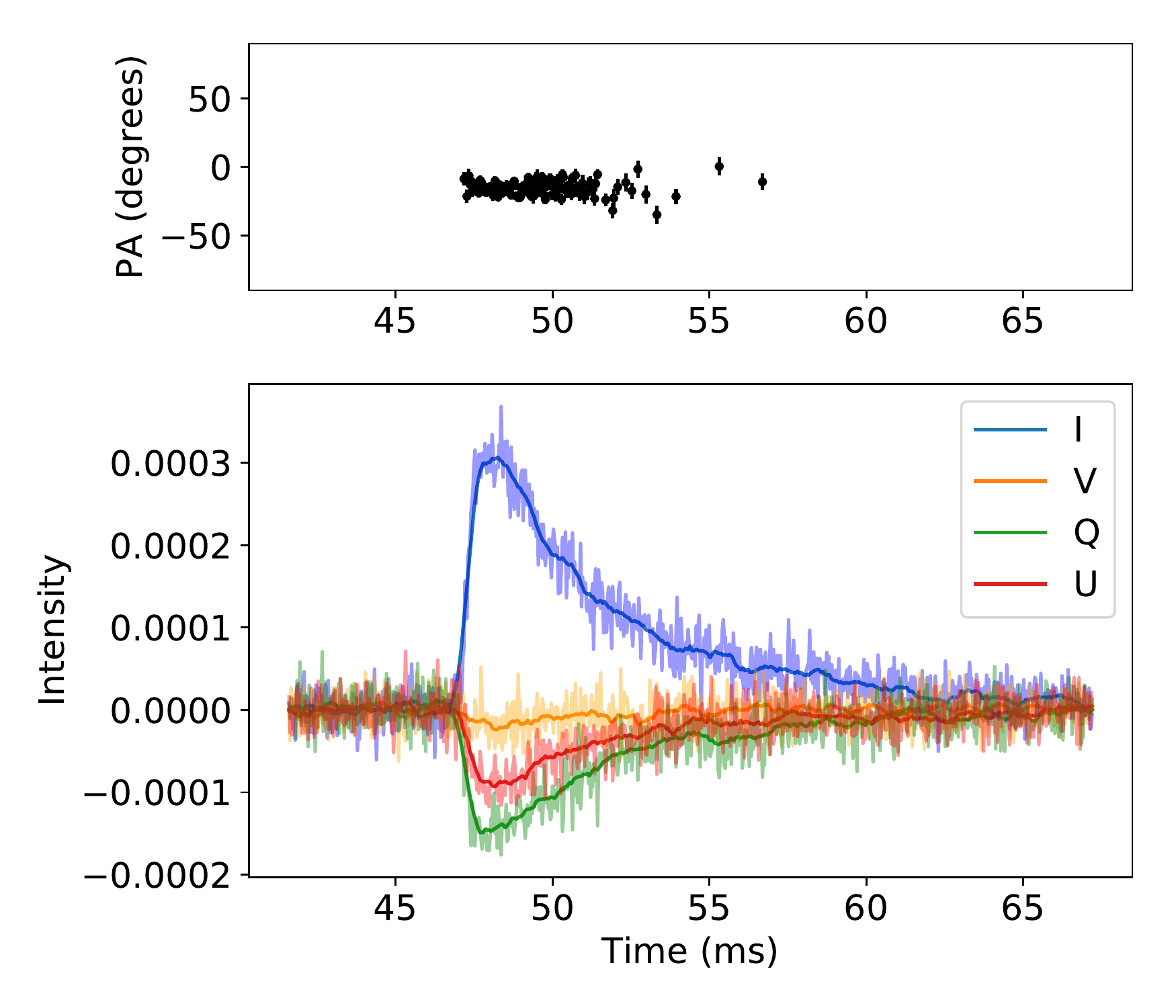}& \hspace*{-0.4in}
\includegraphics[trim={1cm 1cm 5cm 2cm},clip,scale=0.43]{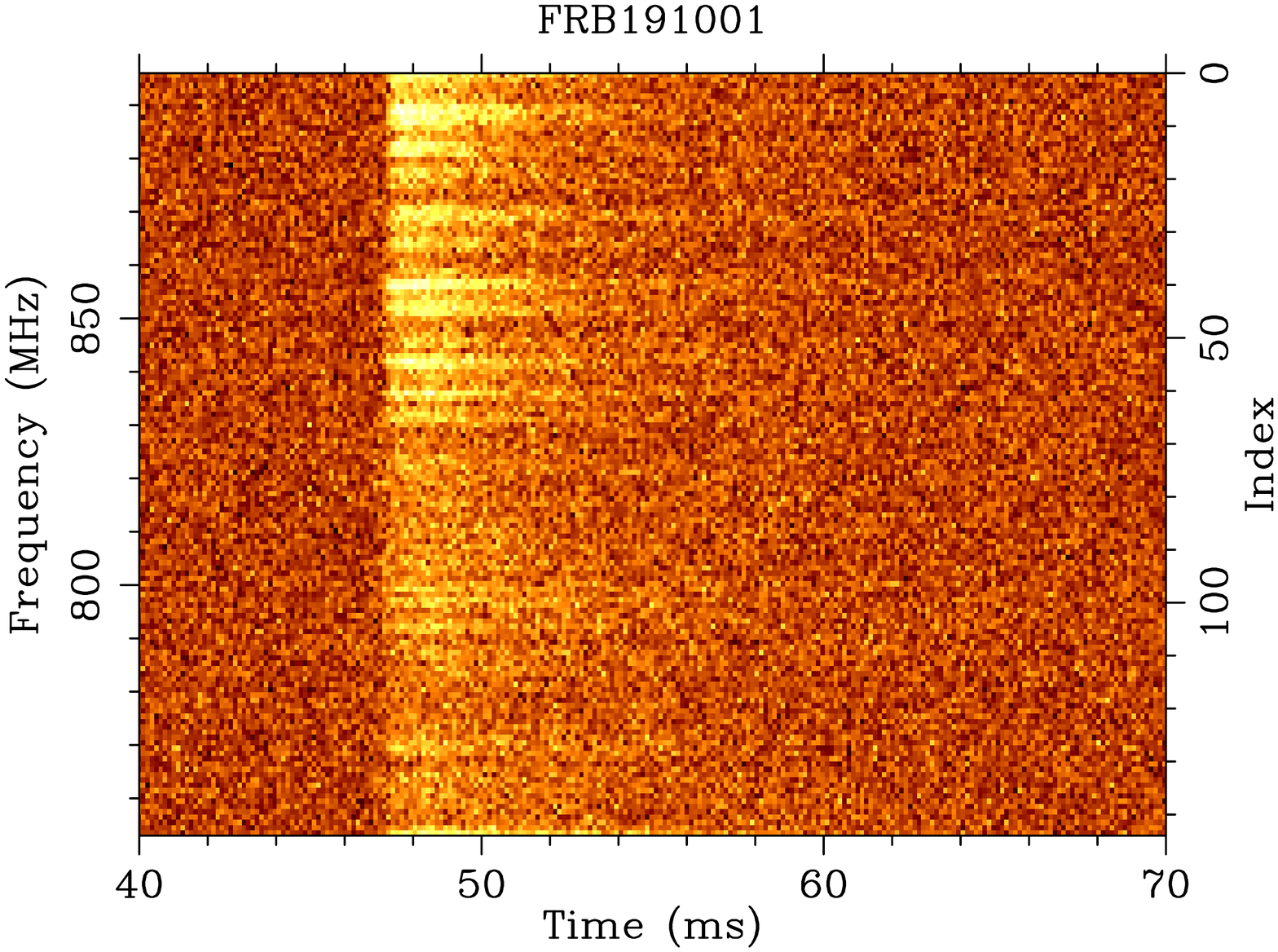} 

\end{tabular}
\caption{Left : The top panel shows the polarization position angle (above $4\sigma$), which is observed to be flat across the majority of the pulse and the bottom panel shows Stokes parameter pulse profiles for FRB~191001 at 824~MHz (see Sec \ref{localisation} for details about the centre frequency). The profiles represent a time resolution of 32~$\upmu$s, and overplotted are $20\times$ smoothed profiles. The FRB shows significant scattering with a scattering time of 3.3~ms at 824~MHz.
Right: The dynamic spectrum of FRB~191001. The data are coherently de-dispersed at a DM of 507.90(7)~pc~cm$^{-3}$. The scintillation bandwidth for the burst is $\sim 600$~kHz at 824~MHz, which is consistent with predictions for diffractive scintillation induced by the Milky Way ISM \citep{Cordes} (see Sec \ref{hightime} for details).}
\label{fig:pulse}
\end{figure*}
\section{Properties of FRB~191001 and its host galaxy}
\subsection{Discovery of FRB 191001}

The burst was detected on 2019 October 1 UT 16:55:35.97081 at a DM of 506.92(4)~pc~cm$^{-3}$, during observations taken as part of the Evolutionary Map of the Universe \citep[EMU;][]{EMU} pilot survey in the frequency range of $752-1088$~MHz. EMU is an ASKAP key science project to conduct a deep radio continuum survey of the entire southern sky. The FRB was detected in beam~24 (an outer ASKAP beam) of the closepack36 beam footprint pattern (see \citet{Shannon2018}) during a $\sim$9~hr observation with 30 antennas in a 336 MHz band centred on 920~MHz. The properties of the burst are presented in Table \ref{tab:properties}. 
Figure \ref{fig:pulse} shows FRB~191001's pulse profile and dynamic spectrum. The discussions of the profile are presented in Section \ref{polarisation}.

\begin{figure*}
\begin{tabular}{cc}
\includegraphics[trim={1cm 13.5cm 0 2cm},clip,scale=0.55]{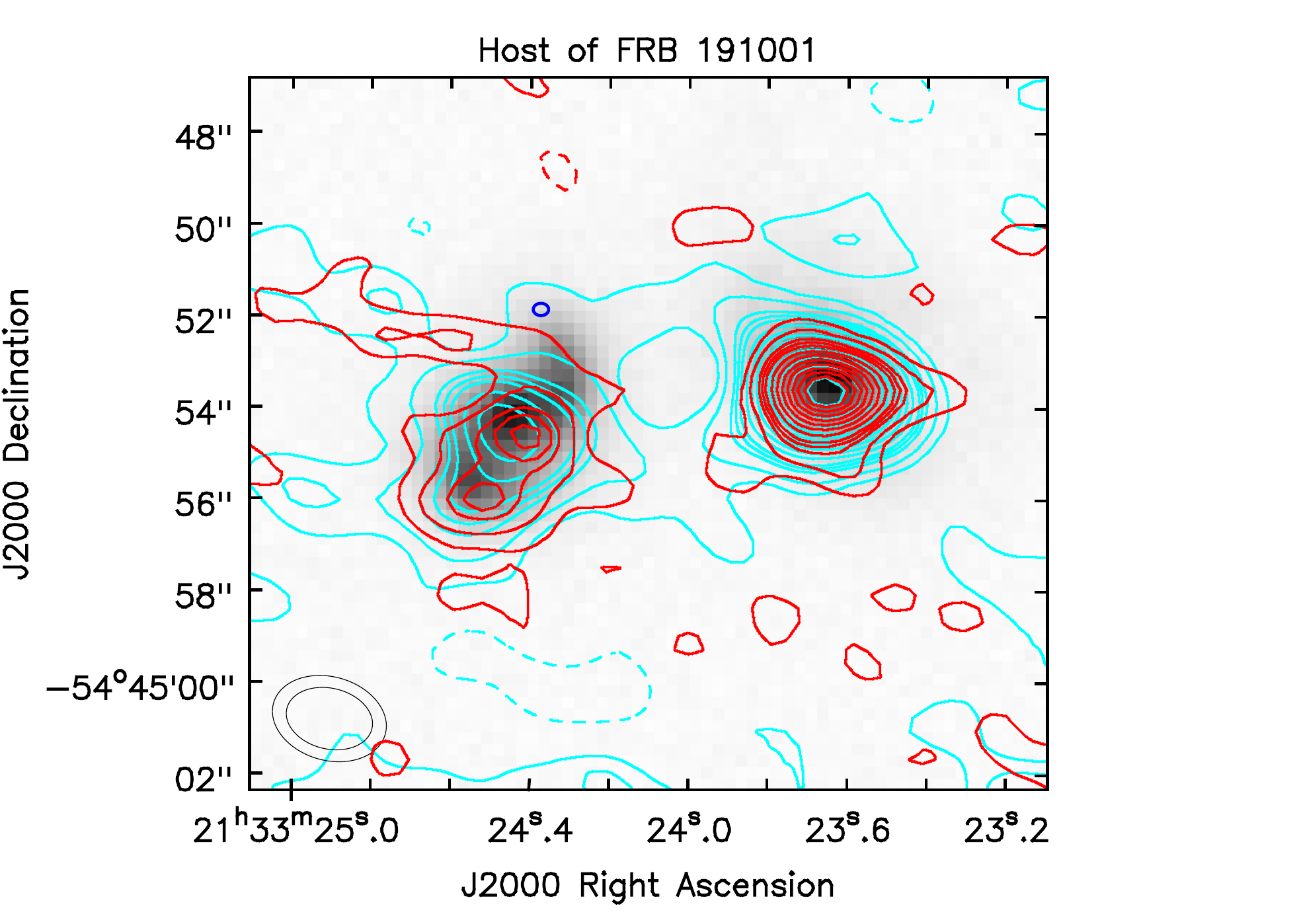}&\hspace*{-0.7in} 
\includegraphics[scale=0.40]{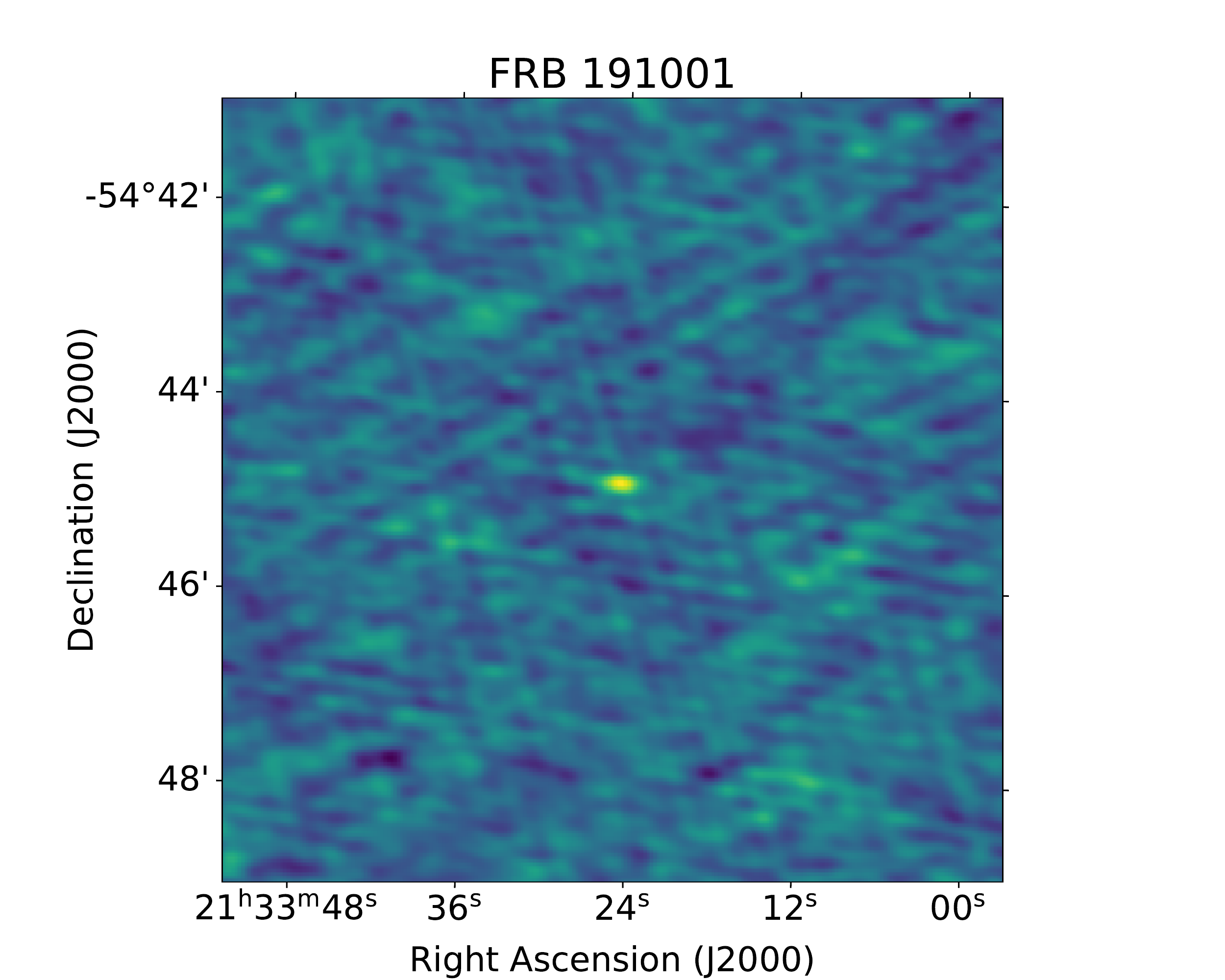} 
\end{tabular}
\caption{
Left: Background $g$-band VLT/FORS2 image of the host of FRB~191001 at a redshift of 0.2340. The blue ellipse marks the 1$\sigma$ uncertainty in the position of FRB~191001 which is 165~mas in RA and 127~mas in Dec. Cyan and red are the contours of radio emission detected with ATCA at 5.5~GHz and 7.5~GHz respectively in levels of [$-$3, 1.5, 3, 5, 6, 7, 9, 11, 13, 15, 17, 19, 21, 25, 30, 35] times the noise of $\sim6~\upmu$Jy beam$^{-1}$ (see Sec \ref{radio properties} for details). Right: Detection of FRB~191001 in the image plane with a 10-s ASKAP snapshot obtained during commensal EMU observations. The FRB is the bright source at the image center with a primary beam-corrected flux density of 19.3~mJy (see Sec \ref{snapshot} for details).}
\label{fig:radio}
\end{figure*}

\subsection{Commensal low time resolution observations}
\label{snapshot}
The standard ASKAP hardware correlator integrates and writes out visibilities on a 10-s timescale --- hereafter termed ``low time resolution" data. The correlator was operating simultaneously, and independently of, the CRAFT FRB search. The low time resolution data were calibrated following the standard data reduction steps as presented in \citet{BhandariVAST} and were then used for imaging.
We performed difference imaging of the consecutive 10-s snapshots around the FRB arrival time derived from the CRAFT data to check for any significant source in the direction of the FRB. FRB~191001 was detected in a difference image providing a preliminary position. Once found, we performed the imaging and deconvolution of that single integration (without differencing) to obtain a better estimate of the flux density. The FRB was detected with 9$\sigma$ significance with a primary beam corrected flux density of 19.3~mJy averaged over 10~s. (see 
right panel of Figure \ref{fig:radio}). We note that the fluence derived from the above flux density is $\sim 1.5\times$ higher than the measured fluence of the FRB. This can be explained by high uncertainty in the measurement of FRB fluence because of an outer beam detection.

\subsection{Localization of FRB~191001}
\label{localisation}
The real-time detection of FRB~191001 in online incoherent-sum data triggered the retention of 3.1~s of raw (voltage) data around the FRB event. Due to a greater-than-usual latency (2.1~s in this case) between the FRB arrival and the voltage dump trigger\footnote{This was caused by a software bug that has since been corrected.}, only half the FRB signal was captured (a center frequency of 824~MHz and bandwidth of 144~MHz), with the upper half of the FRB emission (where the dispersion delay is less) already falling outside the voltage buffer. We followed standard CRAFT procedures for calibrating and imaging both the FRB and background continuum sources \citep{Day2020}, but flagged channels in which the FRB was not present for the FRB image only.

To refine the astrometric position of FRB~191001, we compared the positions of sources identified in the CRAFT image of the FRB field with reference positions obtained from the Australia Telescope Compact Array (ATCA) at 2~GHz, which is the closest we could get compared to ASKAP's FRB observation at 824~MHz. We did not use the full ASKAP observation as a reference due to the known arcsecond level astrometric uncertainties in ASKAP pilot data \citep{BhandariVAST}. The process involved observation of three bright calibrators around the FRB field, namely SUMSS~J212104$-$611125, SUMSS~J213520$-$500652, SUMSS~J220054$-$552008, and three continuum sources detected in the ASKAP image of the field generated from the 3~s of data containing the FRB, namely J2132$-$5420, J2134$-$5450 and J2140$-$5455 at 824~MHz. The phase calibration solutions were derived using each of the calibrator sources and transferred to the background continuum sources and other calibrators. Out of three calibrators, the solutions derived from SUMSS~J213520$-$500652 resulted in zero positional offsets between the known and derived positions (from ATCA image) of calibrators within positional uncertainties. Thus, SUMSS~J213520$-$500652 was used for the remainder of this analysis.

The position of background sources obtained from the ATCA radio image were compared with the ASKAP field source positions, and we obtain a weighted mean systematic offset (also described in \citet{JP+20}) of $731\pm 107$~mas in RA and $-809\pm 99$~mas in Dec. These corrections were applied to the position of the FRB and final uncertainties were obtained by taking a quadrature sum of systematic and statistical uncertainties. The FRB position is RA(J2000): 21:33:24.37(2) and DEC(J2000): $-$54:44:51.86(13).

\subsection{DM excess}
The observed DM of the FRB can be broken down into contributions from various components as
\begin{equation}
    DM_{\rm obs} = DM_{\rm ISM} + DM_{\rm MW,halo} + DM_{\rm cosmic} + DM_{\rm host},
\end{equation}
where $DM_{\rm ISM}$ is estimated to be $44~\rm pc~cm^{-3}$ and $31~\rm pc~cm^{-3}$ from the Galactic models of NE2001 \citep{Cordes} and YMW16 \citep{YMW16}, respectively; the contributions from the Milky Way halo and the host galaxy are $DM_{\rm MW,halo} = 50~\rm pc~cm^{-3}$ and $DM_{\rm host} = 50/(1+z) = 40~\rm pc~cm^{-3}$ respectively, following the assumptions presented in \citet{JP+20}. This leaves the budget for DM from the intergalactic medium (IGM) to be $DM_{\rm cosmic} = 373(386)~\rm pc~cm^{-3}$ using the NE2001(YMW16) models. Based on predictions from the Macquart relation \citep{JP+20}, we would expect $DM_{\rm cosmic}$ to be 203 pc~cm$^{-3}$, which is significantly lower than the value of $DM_{\rm cosmic}$ inferred from the observed DM and assumptions for $DM_{\rm MW,halo}$ and $DM_{\rm host}$. Thus, as for FRB~190608 \citep{Chittidi2020} and FRB~121102 \citep{host}, it is likely that FRB 191001 has a larger host contribution than typical, or lies along a sightline that traces a denser-than-average path through the cosmic web \citep{Simha2020}. The host DM contribution can also be probed by optical studies. We use the relation between optical reddening $E(B-V)$ and hydrogen column density
$N_{\rm H}$ from \citet{Guver2009}, together with the $DM$-$N_{\rm H}$ correlation of \citet{He2013} to estimate the DM contribution from the host of FRB~191001. We find $DM_{\rm host} = 61$~pc~cm$^{-3}$, which is higher than our previous assumption but still leads to an excess in the cosmic DM. A more detailed discussion of the breakdown of the excess DM for this FRB is beyond the scope of this paper, and will be considered in a future work.

\subsection{High time resolution analysis} \label{hightime}
We performed a high-time-resolution analysis of the FRB using the CRAFT voltage data. 
Data were beam-formed (i.e., coherently summed) at the position of the FRB using the delay, bandpass and phase solutions derived from the calibrator source PKS 0407$-$658. The 144 1-MHz-bandwidth ASKAP channels that contained signal from the FRB were then coherently de-dispersed at the FRB DM before being passed through a synthesis filter to reconstruct a single 144-MHz channel with $\sim$7~ns time resolution. A detailed description of the high time resolution construction process is given by \citet{Cho+20}. 

We determined the scintillation bandwidth ($\delta v_{d}$) using a range of time bins in the FRB dynamic spectrum (encompassing roughly the half-power points of the pulse), with frequency resolution of 7.812~kHz, by performing an auto-correlation function (ACF) analysis following \citet{Cho+20}. We estimate $\delta v_{d} \sim 600$~kHz, which is consistent with expectations for diffractive scintillation (DISS) from the Milky Way along the burst line of sight at this frequency using the NE2001 model ($\sim 860$~kHz).

\begin{figure}
\includegraphics[width=0.5\textwidth]{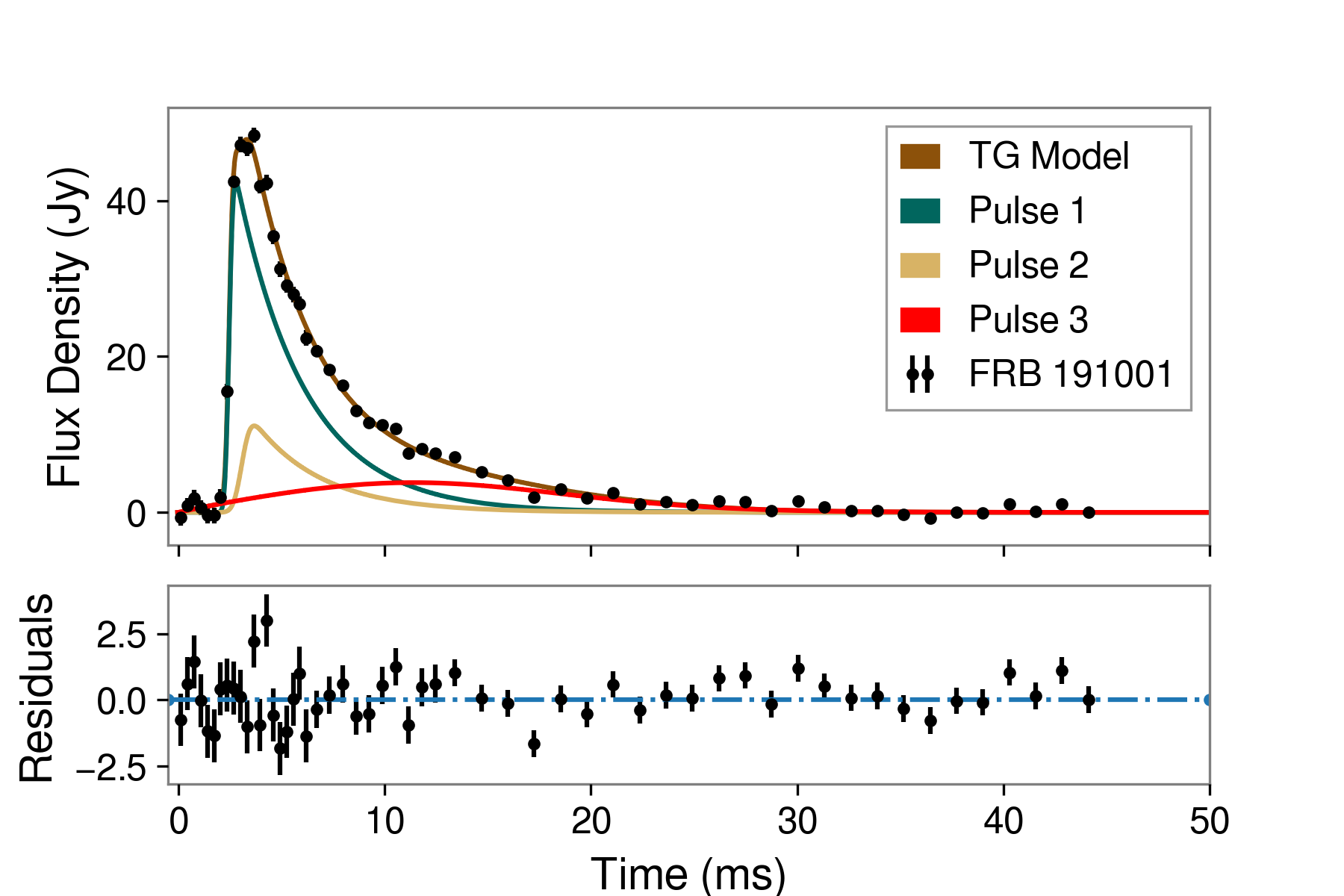} 
\caption{Pulse morphology model fit using 64 $\upmu$s time series. The best fit model comprises three scatter broadened components. 
The residuals are shown in the panel below. The data points between 0--6~ms, 6--25~ms and 25--50~ms are averaged by a factor of 5, 10 and 20 respectively for display.}
\label{fig:gaussian}
\end{figure}

We fit the frequency-averaged pulse profile with scattered Gaussian pulse models using the nested sampling method presented in \citet{Qiu20} and \citet{Cho+20} to compare the evidence for multiple pulse components as demonstrated in \citet{Day2020}. Model comparison favors three scattered pulse components (TG model in Figure \ref{fig:gaussian}) 
by a Bayes Factor of $\rm{log_{10}B= 20}$ to a single pulse model and $\rm{log_{10}B= 5}$ to a double pulse model with significantly lower RMS error. 
The pulse width of components in order of appearance are $0.22\pm0.03$, $0.4\pm0.2$ and $9\pm2$ ms respectively. We measure an exponential broadening of $3.3\pm0.2$~ms at 824 MHz. The second and third pulse component occur $0.6\pm0.2$ and $5.9\pm1.2$~ms after the first pulse (see Figure \ref{fig:gaussian}). 
The presence of further components or frequency-dependent structure could plausibly explain the remaining structure in the residuals in Figure \ref{fig:gaussian}. 

\subsubsection{Spectro-temporal polarization properties}
\label{polarisation}
The time series was time averaged to a resolution of 32~$\upmu$s for polarization analysis (Figure \ref{fig:pulse}). The polarization calibration was performed by comparing an ASKAP observation of the Vela pulsar with an observation obtained with the Parkes radio telescope for which an accurate polarization calibration exists, as described in \citet{Day2020}. Unlike some earlier FRBs detected by ASKAP, there is negligible leakage
between Stokes $U$ and $V$ as a result of polarization leakage correction in the beam weight calculation \footnote{\url{https://www.atnf.csiro.au/projects/askap/ASKAP_com_update_v32.pdf}}.

We used the \texttt{RMFIT} program in the \texttt{PSRCHIVE} software package \citep{PSRCHIVE} to calculate the rotation measure (RM) of FRB~191001. \texttt{RMFIT} performs a search for peak linear polarization as a function of trial RM. We find the best fit RM for FRB~191001 to be $55.5\pm0.9$~rad~m$^{-2}$. 

In addition, we estimated polarization fractions integrated over the FRB pulse using the \texttt{PSRCHIVE} software package. The values for total, linear and circular polarization fractions are presented in Table \ref{tab:properties}.
These values are lower than the values observed for the sample of ASKAP FRBs discussed in \citet{Day2020} and \citet{Cho+20}.
We also note that the circular (Stokes V) peaks later than the linear polarization.

The left panel of Figure~\ref{fig:pulse} shows the Stokes parameter profiles and polarization position angle (PA) for FRB~191001, with a time resolution of 32~$\upmu$s. As noted above, the pulse is not consistent with a single Gaussian component convolved with an exponential scattering tail -- indicating the presence of multiple components blended in the scattering tail (as proposed in \citealp[][]{Day2020} for FRB~180924 and FRB~190608). We observe a pulse rise time of $\sim 640\, \upmu$s by counting the number of samples from 1$\sigma$ to the peak value. Also, the PA is initially flat across the majority of the pulse (as shown in the top left panel of Figure \ref{fig:pulse}), which is possibly a consequence of scattering that not only affects the shape of the total intensity pulse profile but also measured polarization properties \citep{Caleb2018}. At later times, the noise in the PA makes it difficult to establish whether the second component has a position angle that differs substantially from the scattering tail of the first component. It is suggestive that the largest deviations from a constant PA are seen at this time.

The absence of variation in the PA across the main pulse and the scatter-blended multiple components observed in FRB~191001 bear a remarkable resemblance to FRB~180924 and FRB~190608 \citep{Day2020}. The separation between the three modeled components for FRB~191001, however, greatly exceeds that of the other bursts. Of note, although observed at different frequencies, the three FRBs exhibit similar substructure in the residuals in the scattering tail.

\subsection{Host galaxy of FRB~191001}

The host galaxy of FRB~191001 was identified as
DES~J213324.44$-$544454.18, a galaxy catalogued in the Dark Energy Survey \citep[DES]{DES}. Spectroscopic observations conducted on 2019 October 4~UT using the Gemini Multi-Object Spectrograph (GMOS) on the Gemini-South telescope established the redshift of the host to be $z=0.2340(1)$ using the {H$\beta$} spectral line \citep[see][for further details]{heintz2020}. 
On 2019 October 05~UT deep optical imaging observations were performed with the FOcal Reducer and low dispersion Spectrograph 2 (FORS2) instrument on the Very Large Telescope (VLT). The left panel of Figure \ref{fig:radio} shows the $g$-band image of the host, which is clearly a spiral galaxy. The neighbouring galaxy, DES~J213323.65$-$544453.6, is also at a similar redshift ($z=0.2339(2)$), hence the system is a double and possibly interacting pair of galaxies.
Detailed optical properties of the host are described in \cite{heintz2020}.  

\begin{figure}
\includegraphics[scale=0.4]{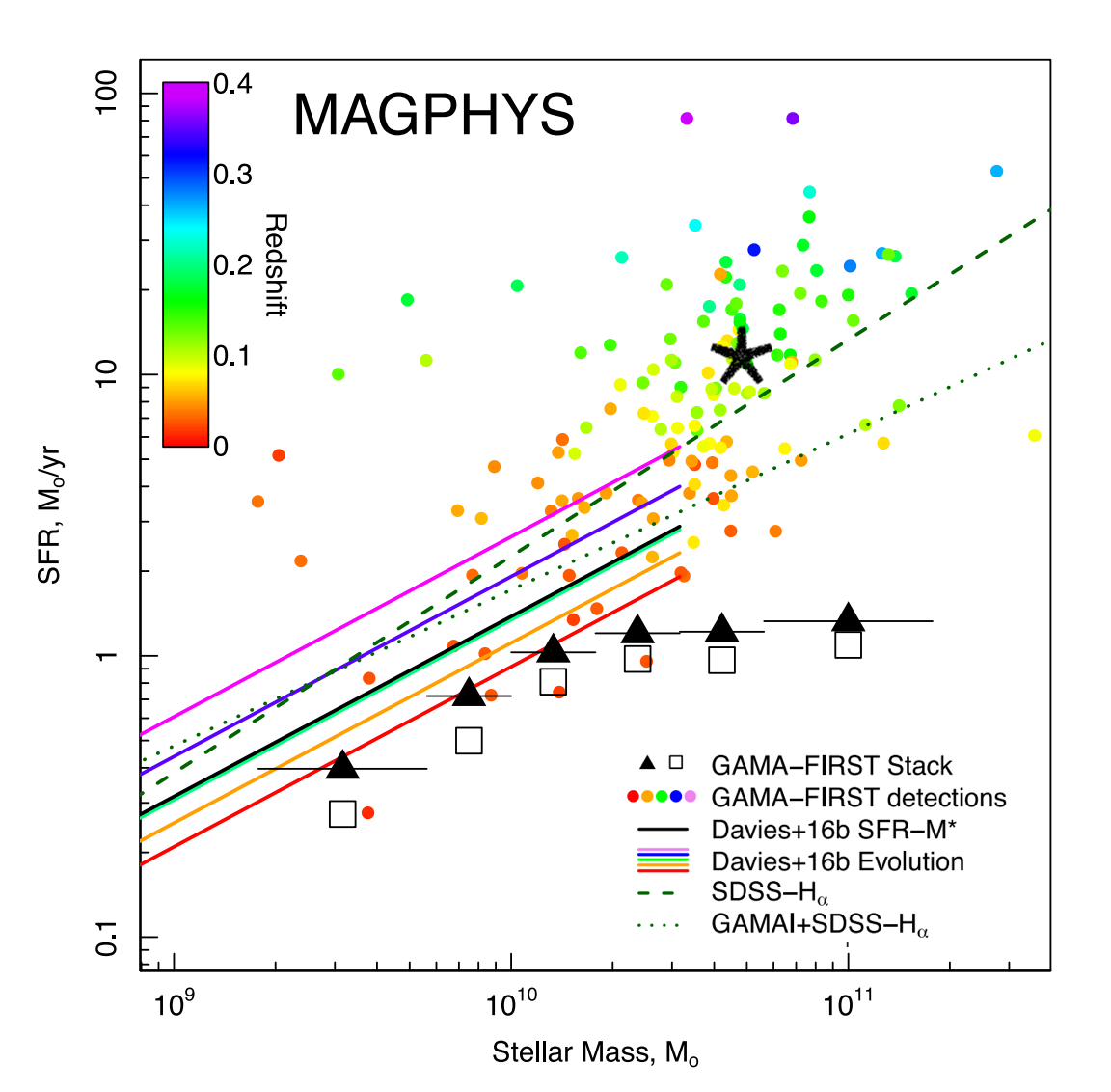} 
\caption{The 1.4 GHz SFR$-$M\text{*} relation. We show the relation derived from GAMA galaxies, derived using free-fit luminosity-to-SFR relation for MAGPHYS. The black star is the host of FRB~191001 overplotted on a sample of galaxies at various redshifts detected in the GAMA survey. Figure adapted from Figure 5 of \citet{Davis2017} }
\label{fig:sfr}
\end{figure}
\begin{table}
\centering
    \begin{tabular}{cccc}
    \hline
 & & Host   & Western Source     \\ 
\hline
   Freq.  & rms & S$_{\rm int}$ & S$_{\rm int}$ \\
   (GHz) &($\upmu$Jy~beam$^{-1}$) &($\upmu$Jy)  & ($\upmu$Jy)   \\
   \hline
 
   2 & 30 & $241 \pm2$ & $509 \pm 2$\\
   5.5 & 11& $103\pm 1$ & $246 \pm2$ \\
   7.5  &10 & $123\pm2$ &  $208\pm1$ \\
    \hline
    \end{tabular}
\caption{Integrated flux densities for the host of FRB~191001 and neighbouring western source derived from epoch 4. The flux densities and uncertainties were estimated using \texttt{imfit} in \texttt{miriad} at 5.5 and 7.5~GHz images, smoothed to a common resolution ($4.7^{''}\times3.4^{''}$, PA $= -1.8$ deg) }
\label{tab:atca}
\end{table}
\subsection{Radio properties of the Host galaxy}
\label{radio properties}
We conducted observations with ATCA (project code C3211) in a 6-km array configuration to search for radio emission from the FRB\,191001 host galaxy at 5.5~GHz and 7.5~GHz. Observations were performed in three epochs starting 2020 January 24, 2020 March 06 and 2020 March 12 at different hour angles to maximise ($u,v$)-coverage. Epoch~1 was badly affected by weather and therefore discarded. We later obtained a fourth epoch on 2020 April 16 (project code C3347) at 2~GHz, 5.5~GHz and 7.5~GHz. We combined epoch~2, epoch~3 and epoch~4 data in the 4~cm band for deep imaging. However, we use only epoch~4 data for estimating source flux densities. 

\subsubsection{Search for persistent emission}
We used the obtained ATCA data to search 
for a compact persistent radio source that may be associated with FRB~191001. We detect low level diffuse radio emission with a peak flux density of $\sim 15~\upmu$Jy~beam$^{-1}$ at the FRB position pixel, corresponding to a luminosity of $2.1\times 10^{21}$~W~Hz$^{-1}$ at 5.5~GHz. This luminosity is an order of magnitude less than the luminosity of persistent source observed for FRB~121102 \citep{VLAlocalisation}. We did not find a compact persistent radio source co-located with FRB~191001 above a flux density of $15\upmu$Jy.

\subsubsection{Star-formation in the host galaxy}
We detect radio emission from the host of FRB~191001 (H) and the western source (W) as shown in the left panel of Figure \ref{fig:radio}. 
For estimating the integrated flux densities, we smoothed the images at respective frequencies to the same angular resolution -- smoothing beam resolution at 7.5~GHz ($1.7^{''} \times1.2^{''}$, PA $= 2.1$ deg) and 5.5~GHz ($2.3^{''} \times 1.7^{''}$, PA $= 2.2$ deg) to 2~GHz ($4.7^{''} \times 3.4^{''}$, PA $= -1.8$ deg) resolution. The flux densities for both sources were measured using the \texttt{miriad} task \texttt{imfit} on the smoothed version of radio images and fixing the beam to the size of the common resolution, i.e.,~$4.7^{''} \times 3.4^{''}$, PA $= -1.8$ deg. Flux densities are presented in Table \ref{tab:atca}. We note an excess emission at 7.5~GHz to the south-east of the host galaxy. This resolved flat-spectrum component overlaps with the south-east spiral arm of the galaxy (See Figure \ref{fig:radio}) could indicate the presence of thermal emission associated with (possibly ongoing) star-formation along the spiral arm. Further discussion will be presented in an upcoming paper. 

Next, a spectral index map was obtained using the same resolution images at 2~GHz and 5.5~GHz. 
We find the spectral index ($\alpha$), where $S_\nu \propto \nu^\alpha$, for the host 
vary from $-1.0$ at the outer edges to $-0.8$ at the center. Hence, the lack of much flattening ($\alpha>-0.7$) of the spectral index near the nucleus suggests no evidence for a dominant compact AGN. We further fit a power law to the integrated flux densities (excluding 7.5~GHz for the host) and find their spectral indices to be $\alpha_{H} = -0.8$ and $\alpha_{W} = -0.7$, respectively.
The diffuse morphology and steep negative spectral index suggest that the radio emission (or most of the emission seen in the host) is dominated by synchrotron emission due to star-formation (SF) in the galaxy. We note that the current resolution observations do not rule out contamination from a low level AGN emission.

Furthermore, we estimate an inferred SFR using a new 1.4~GHz luminosity-to-SFR relation derived combining the data from the Galaxy And Mass Assembly (GAMA) survey and the Faint Images of the Radio Sky at Twenty-cm (FIRST) survey \citep{Davis2017}. The new robust calibration to the 1.4 GHz$-$SFR relation is given by
\begin{equation}
   \rm  log_{10}[\rm SFR(M_\odot\ yr^{-1})] = 
M \times log_{10}[\rm L_{1.4~GHz}(W~Hz^{-1})] + C
\end{equation}
where parameters, M $= 0.75 \pm 0.03$ and C $= -15.96\pm0.58$ for correction based on the \texttt{MAGPHYS} SFRs in GAMA work and $\rm L_{1.4}$ is the radio luminosity at 1.4~GHz. We find the star formation rate of the host to be SFR$_{\rm MAGP} = 11.2~\rm M_\odot\ \rm yr^{-1}$ and that the host of FRB~191001 is consistent with the underlying population of galaxies at a similar redshift in the GAMA sample (see Figure \ref{fig:sfr}).
A possible caveat is the potential for contamination of a low level AGN in our star formation rate calculations. Assuming the source is dominated by SF, gives us an upper limit on the SFR. 

A high SFR qualifies this galaxy as starburst or experiencing interaction triggered bursts of star formation similar to the NGC4038/4039 pair of galaxies a.k.a The Antennae \citep{Whitemore}. We note that the host of FRB~191001 has the highest SFR in the ASKAP sample \citep{heintz2020}, suggesting diversity in galaxies hosting FRBs. 

\begin{figure*}
\begin{tabular}{cc}
\includegraphics[scale=0.255]{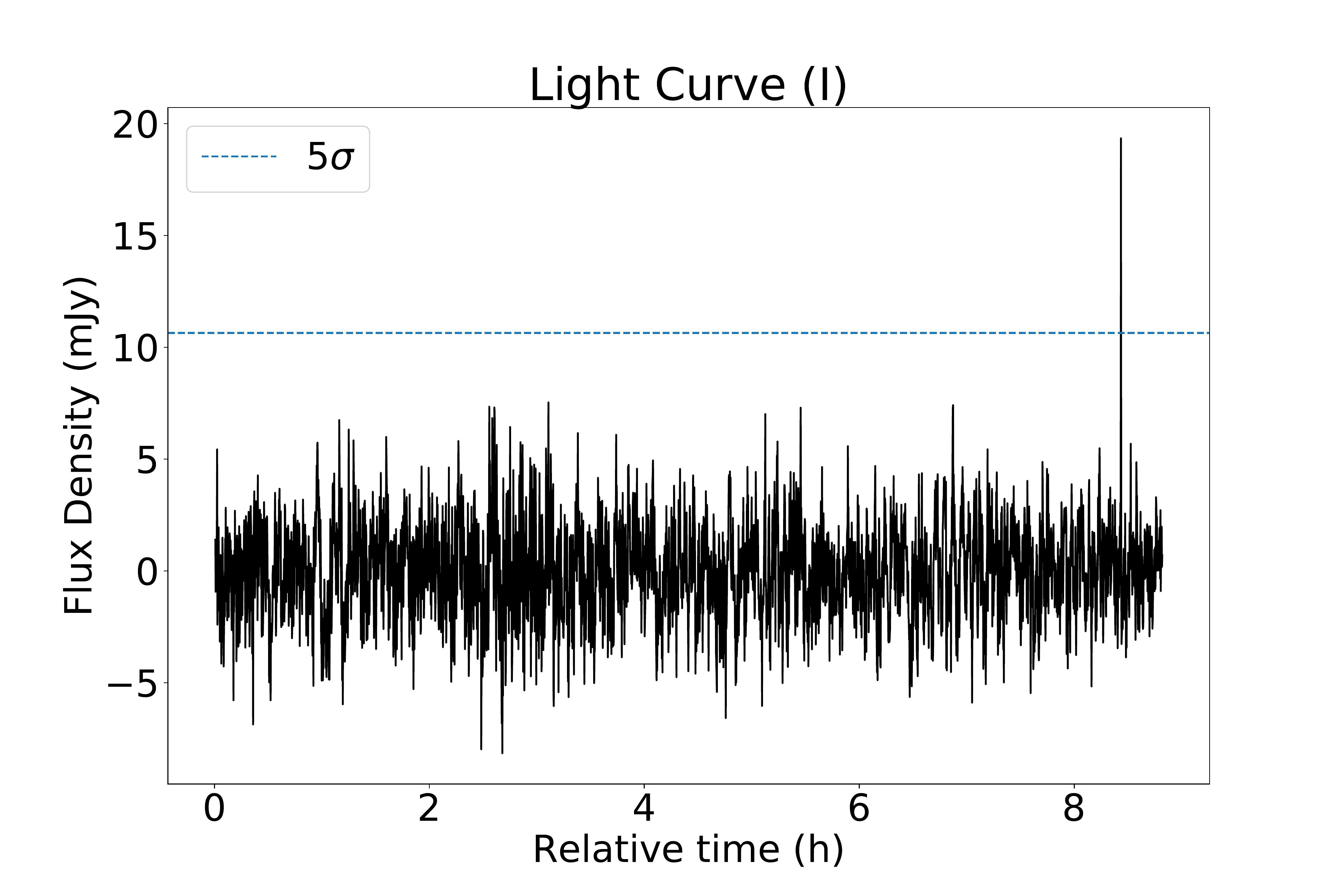} \hspace*{-0.3in}
\includegraphics[scale=0.25]{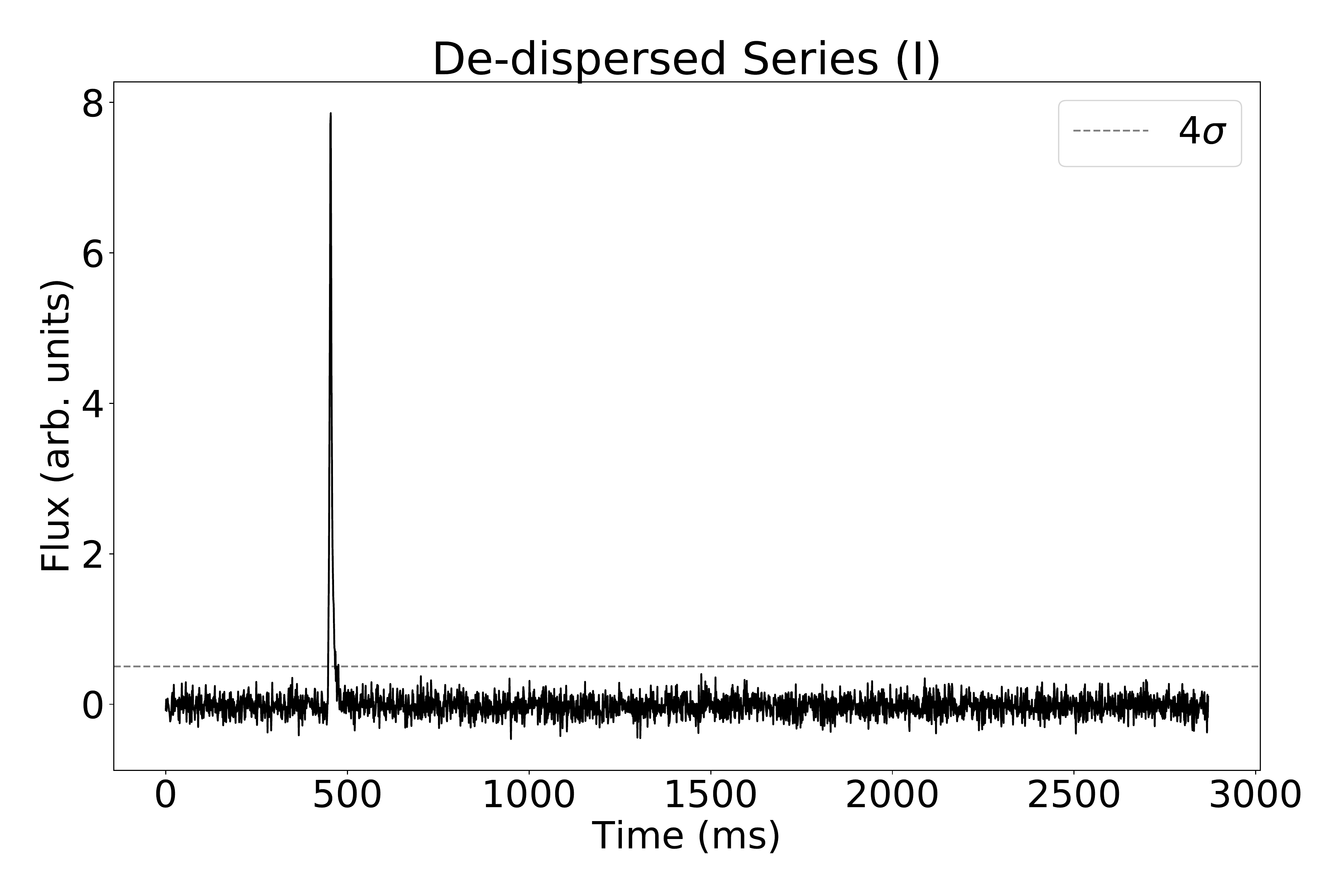} 
\end{tabular}
\caption{
 Left: ASKAP lightcurve obtained  at the position of the FRB on a timescale of 10~s for an $\sim$9~hr observation of the EMU field. FRB~191001 was detected at UT 16:55:35.97081 in ASKAP low time resolution data. Right: ASKAP lightcurve on a timescale of 1~ms derived by beamforming CRAFT data, de-dispersed at the DM of the FRB.
}
\label{fig:timeseries}
\end{figure*}

\section{Search for Radio afterglows}
\subsection{Low time resolution data}
We performed a search for repeats and afterglows at the position of FRB~191001. Firstly, a deep image of the field was made using ASKAP low time resolution data taken $\sim 8$~hrs prior to the FRB observation. A model of background sources was created using the \texttt{CASA} task \texttt{TCLEAN} and subtracted from all ASKAP data using \texttt{UVSUB}. Secondly, the visibilities were rotated in phase to the position of the FRB to extract the dynamic spectra. Finally, we obtained a light curve for $\sim$9~hrs of data (see left panel of Figure \ref{fig:timeseries}) by averaging baselines of the source-model-subtracted and phase-rotated data. We also subtracted the underlying host radio emission from the light curve. The peak in the light curve (Stokes I) is the emission from the FRB. 

\begin{figure}
\includegraphics[scale=0.4]{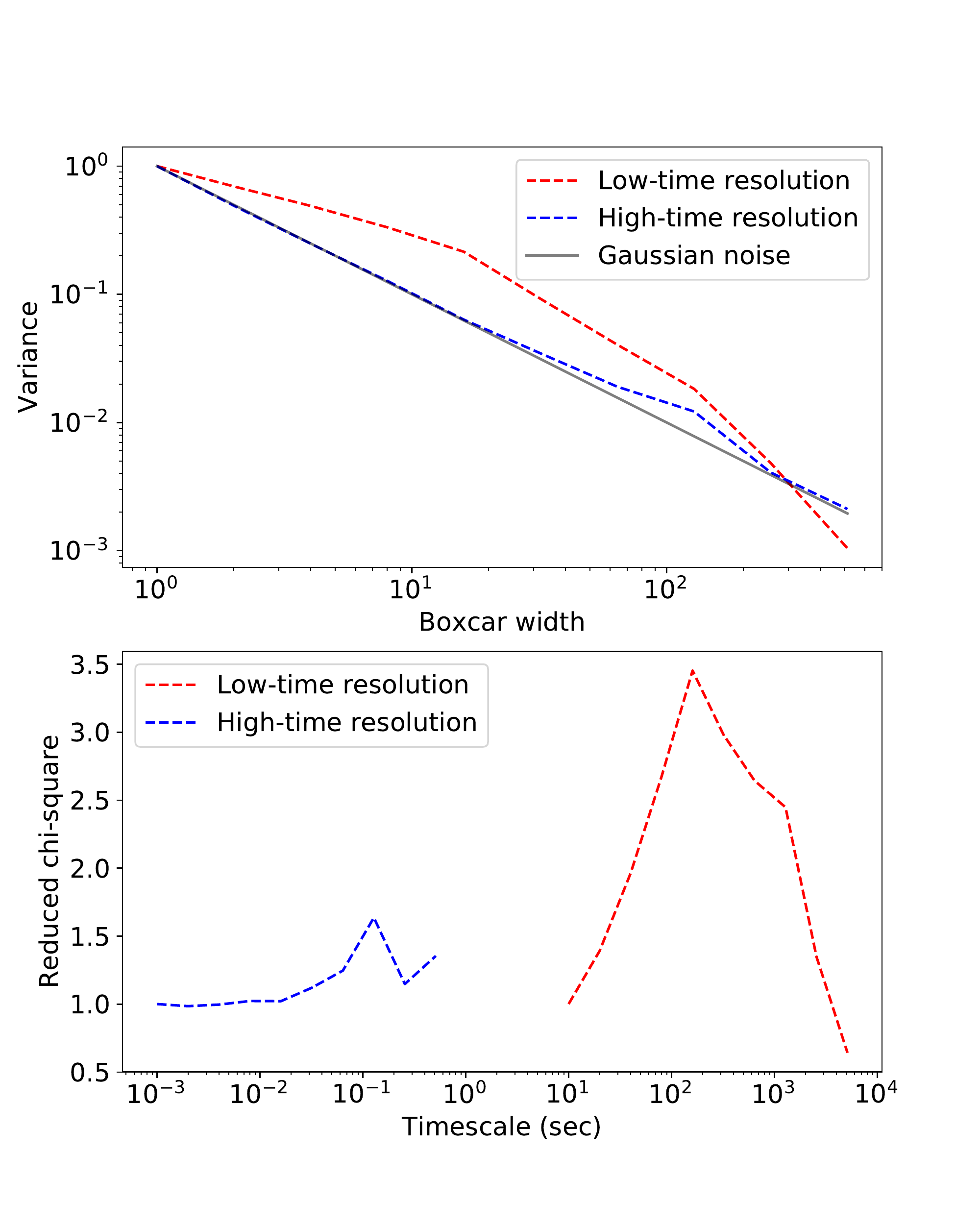}
\caption{Top Panel: Plot of variance versus boxcar width for ASKAP low time resolution series (red) and CRAFT high time resolution series (blue) after subtracting FRB~191001. This plot shows that the Gaussian noise in low time resolution series data does not vary linearly with boxcar width (as expected for white Gaussian noise). Bottom Panel: Plot of reduced chi-square vs. timescales under investigation assuming a white Gaussian noise. No varying radio emission was observed in low time resolution data lightcurve above 5$\sigma$. The high value of reduced chi-square is due to systematics. A chi-square test on high time resolution series data rejects the hypothesis that varying radio emission exists at 98\% confidence.}
\label{fig:stats}
\end{figure}

We averaged the light curve over boxcar widths of $2^{N}$, where $N$ varies from $0-9$, to search for slowly varying radio emission on timescales ranging from 10~s to 1.4~hr. The data shows red noise either due to contamination by sidelobes of poorly subtracted faint sources in the image domain, radio frequency interference (RFI), or residual emission from the host of FRB~191001 (see top panel of Figure \ref{fig:stats}), and is therefore not amenable to a chi-square test that assumes statistically independent (white) samples.
By eye, we find no evidence for slowly varying radio emission at the FRB position $\sim8$~hr before and $\sim1$~hr after FRB~191001 above a $5\sigma$ flux density limit at respective timescales as presented in Table \ref{tab:limits}.
\begin{table}
    \centering
    \begin{tabular}{ccc}
    \hline
   Timescale & Flux-limit & luminosity \\
    (sec) & (mJy beam$^{-1}$) & (W Hz$^{-1}$)\\
    \hline
10  & 10.64 & 1.50 $\times 10^{24}$\\
20  & 8.87 & 1.25$\times 10^{24}$ \\
40 & 7.44 & 1.05$\times 10^{24}$ \\
80 & 6.14 & 8.67$\times 10^{23}$ \\
160 & 4.93 & 6.95$\times 10^{23}$ \\
320 & 3.23 & 4.56$\times 10^{23}$ \\
640 & 2.14 & 3.02$\times 10^{23}$ \\
1280 & 1.44 & 2.03$\times 10^{23}$ \\
2560 &0.74 & 1.04$\times 10^{23}$ \\
5120 & 0.34 & 4.85$\times 10^{22}$ \\
   
    \hline

    \end{tabular}
    \caption{$5\sigma$ flux density and limits on luminosities for radio emission at different timescales at the position of the FRB. }
    \label{tab:limits}
\end{table}
\subsection{High time resolution data}
We also performed a search for dispersed radio emission pre/post-FRB in $\sim$3~s of CRAFT high time resolution data (presented in Section \ref{hightime}). We time-averaged the beam-formed data to 1~ms (see right panel of Figure~\ref{fig:timeseries}) to obtain an initial time series, which was further averaged over boxcar widths $2^{N}$, where $N$ varies from $0-9$. This allowed a search for varying radio emission on timescales ranging from 1~ms -- 512~ms. Analysis of the noise showed it to be Gaussian-distributed with no frequency dependence (i.e., white noise). Therefore, we performed a chi-square analysis to test the null hypothesis that there is a slow varying radio emission pre/post-FRB. The measured chi-square ($\chi^{2}_{\rm M}$) is given by

\begin{equation}
    \chi^{2}_{\rm M} = \displaystyle\sum_{i=0}^{i=n} \frac{(Y_{i} - Y_{\rm mean})^{2}}{\sigma_{i}^{2}} 
\end{equation}
where $Y_{i}$ is the flux density for boxcar width $i$, $Y_{\rm mean}$ is the mean flux density of the time series and $\sigma_{i}$ is the standard deviation of the off-pulse time series scaled as $\sigma_{i} = \sigma_{0}/ \sqrt{i}$ ($\sigma_{0}$ is the standard deviation of the subtracted FRB time series with zero boxcar width and $i = 2^{N}$, where $N$ varies from $0-9$). We calculated the cumulative distribution function (CDF) and the probability $P$ of obtaining a measured chi-square ($\chi^{2}_{\rm M}$) by chance and compared it with the critical chi-square ($\chi^{2}_{\rm crit}$) for $N$ degrees of freedom. Variable radio emission exists if $\chi^{2}_{\rm M} > \chi^{2}_{\rm crit}$ for $P < 0.02$ (98\% confidence level). In our data, the $\chi^{2}_{\rm M}$ for each boxcar width was less than $\chi^{2}_{\rm crit}$. Therefore, 
we rejected the null hypothesis that varying radio emission exists at the 98\% confidence level.

\section{Discussion}

\subsection{FRBs as image plane transients}
Here we consider the viability of detecting FRBs as image-plane transient in 10-s snapshot images with ASKAP. A $5\sigma$ detection in a 10-s ASKAP snapshot would be very difficult to confirm. We use $10\sigma$ or flux density greater than 20~mJy (robust=0.5 weighting) as a reasonable threshold, which should lead to minimal false positives.
 Of the CRAFT sample of 25 FRBs discovered in fly's-eye mode \citep{Shannon2018, Bhandari+19, Qiu+19}, 3 have flux densities greater than 20~mJy, when diluted to 10~s integration time. Thus, three out of 25 FRBs would have been detected as reliable image plane transients. Scaling the ASKAP all-sky rate, R($>26$~Jy~ms $(\rm w/1.26~\rm ms)^{-1/2})$ of  37~sky$^{-1}$~day$^{-1}$ \citep{Shannon2018}, we derive an FRB rate, $R (>20~\rm mJy, 10~\rm s)$ of $4.4$ FRBs~sky$^{-1}$~day$^{-1}$ or
$0.00011$~deg$^{-2}$ day$^{-1}$.
We compare this rate with other slow transients in Table \ref{tab:transients}. A scaling of $N \propto S^{-3/2}$ has been used to scale the rate of various slow transients above a flux density threshold of 0.3~mJy in \citet{mhb+16} to the ASKAP flux density limit of 20~mJy. The FRB transients are more common than all other slow transients except the AGN(ISS) and the only transients expected on 10-s timescale. 

\textit{How useful is a 10-s search of ASKAP data?} The 10-s data search is affordable as compared to searching at different DM trials. However, DM searching would not be possible to confirm candidates as the dispersion delay is typically $\ll 10$~s for FRBs less than a DM of $\sim2000~\rm pc~cm^{-3}$ at 920 MHz. 
Additionally, the existing incoherent sum (ICS) CRAFT search system can find FRBs more effectively. The S/N in the 10-s data is improved by $\sqrt{N}$, where $N$ is the number of antennas, but diluted by $\sqrt{\rm (10~s/FRB\_width)}$. For instance, a $100\sigma$ FRB of width $1$~ms in ICS mode with $N=36$ antennas will be a $6\sigma$ detection in ASKAP low-time-resolution imaging mode; hence ICS mode is significantly more sensitive than the ASKAP imaging mode. Nevertheless, we encourage searches for transients at shorter timescales 
at other interferometric sites which lack CRAFT-like system for potential FRB detection.

\begin{table}
    \centering
    \begin{tabular}{ccc}
    \hline
    Object &  Timescale & Rate ($>$20~mJy)  \\
    & & (deg$^{-2}$) \\
    \hline
    AGN (ISS) & mins$-$days & 0.110 \\
    Active star (flaring) & hrs$-$days & $3.7 \times 10^{-5}$  \\
    Active binary (flaring) & hrs$-$days & $1.8 \times 10^{-5}$ \\
    CV (Dwarf nova/jet) & hrs$-$days & $1.8 \times 10^{-6}$ \\
    YSO (mass accretion/flare) & hrs$-$weeks & $< 9.1 \times 10^{-5}$ \\
    Brown Dwarfs (pulsing) & sec$-$hrs & $< 9.1 \times 10^{-5}$ \\
      \hline
    \end{tabular}

    \caption{Rates of the slow radio transients at ASKAP snapshot sensitivity modified from \citet{mhb+16}. The slow transient sources include interstellar scintillation (ISS) of an AGN, active star flaring, cataclysmic variables (CV), young stellar object (YSO) and brown dwarfs. These transients may vary on timescales of seconds to weeks.}
        \label{tab:transients}
\end{table}

\subsection{Constraints on afterglows}

\citet{WangLai2020} have predicted radio afterglows for models involving a NS, WD or BH as the central engines, finding that sub-mJy level afterglows for non-repeating FRBs peak around 10 days post-burst at 1~GHz. For a binary NS merger scenario, the $\upmu$Jy level peak of the radio afterglow light curves of the jet (at different viewing angles) and isotropic ejecta at 1.4~GHz from a source at $z=0.5$ may vary on time scales of a few days to years \citep{Lin2020}. For a magnetar produced in a binary NS merger and AIC of a white dwarf, a late-time radio emission (from months/years to decades depending on the triggering mechanism) is anticipated \citep{Margalit2019}.

Most of these models predict FRB afterglows on timescales of months to years, 
longer than our observations probe. 
More interestingly for our study, \citet{Yi2014} present light curves for both forward- and reverse-shock afterglows on timescales of seconds to days post-burst. 
They used the standard fireball model for GRB afterglows to estimate luminosities of FRB radio afterglows for a range of assumed total kinetic energies and redshifts. Their models predict radio emission with post-burst flux densities $< 1$~mJy for burst redshifts between 0.01 to 0.5 at 1~GHz. Our 920 MHz ASKAP data probe these timescales but the predicted fluxes are well below our search threshold of 10~mJy for a $5\sigma$ detection of radio emission from an FRB at $z=0.2340(1)$, consistent with our non-detection of afterglows in low time resolution data spanning the hour after the FRB.

Our luminosity limits are presented in Table \ref{tab:limits}. Detection of energetic and low redshift FRBs ($z = 0.1-0.01$) in commensal ASKAP-CRAFT observations will place stronger constraints on the radio radiative efficiency of this model or could lead to detection. Constraints on long-lasting, persistent or variable radio emission associated with FRBs will require a long term monitoring program of FRB host galaxies on a day to year timescales.

\section{Summary}
We report the detection of the first commensal FRB with ASKAP, FRB~191001, at 920~MHz. Simultaneous imaging with the ASKAP hardware correlator led to a search for slowly varying radio emissions before, during and after the FRB. We did not find a varying radio emission and report luminosity limits on timescales from 10~s to 1.4~hrs, which could potentially be used to constrain progenitor models predicting FRB afterglows. We also demonstrate that bright FRBs can be detected as image-plane transients.

FRB~191001 is the brightest burst which has the best localisation among the sample of seven ASKAP localised FRBs \citep{JP+20}. The FRB originates from the outskirts of a star-forming spiral galaxy in a possibly interacting double system at a redshift of $0.2340(1)$. Radio observations of the host galaxy reveal no compact persistent radio source associated with FRB~191001 above 15$\upmu$Jy. However, the host is radio luminous with most of the synchrotron radio emission occurring due to high star-formation in the galaxy. 

FRB 191001 shows multiple burst components, a large scattering
tail and a flat polarisation position angle. These properties bear similarities with FRB~180924 and FRB~190608 \citep{Day2020}. While the FRB is hosted in a star-forming galaxy, the low Faraday rotation hints at a progenitor environment not dominated by high magnetic fields.

The commensal observations of ASKAP-CRAFT with the imaging mode will continue to explore prompt radio emissions and afterglows associated with FRBs (if any), making ASKAP a unique and a powerful instrument in the studies of FRBs and their progenitor systems.

\section*{Acknowledgements}
SB would like to thank Manisha Caleb for discussions about polarization of FRB~191001. Based on observations collected at the European Southern Observatory under ESO programme 0103.A-0101(A). K.W.B., J.P.M, and R.M.S. acknowledge Australian Research Council (ARC) grant DP180100857.
A.T.D. is the recipient of an ARC Future Fellowship (FT150100415).
R.M.S. is also the recipient of an ARC Future Fellowship (FT190100155). J.X.P., as a member of The Fast and Fortunate for FRB Follow-up team, acknowledges support from NSF grants AST-1911140 and AST-1910471. NT acknowledges support by FONDECYT grant 11191217. The Australian Square Kilometre Array Pathfinder is a part of the Australia Telescope National Facility which is managed by CSIRO. Operation of ASKAP is funded by the Australian Government with support from the National Collaborative Research Infrastructure Strategy. ASKAP uses the resources of the Pawsey Supercomputing Centre. Establishment of ASKAP, the Murchison Radio-astronomy Observatory and the Pawsey Supercomputing Centre are initiatives of the Australian Government, with support from the Government of Western Australia and the Science and Industry Endowment Fund. 
We acknowledge the Wajarri Yamatji as the traditional owners of the Murchison Radio-astronomy Observatory site. 
The Australia Telescope Compact Array is part of the Australia Telescope National Facility which is funded by the Australian Government for operation as a National Facility managed by CSIRO. We acknowledge the Gomeroi people as the traditional owners of the Observatory site.
The Gemini-S/GMOS observations were carried out as part of program GS-2019B-Q-132, obtained at the international Gemini Observatory, a program of NSF’s OIR Lab, which is managed by the Association of Universities for Research in Astronomy (AURA) under a cooperative agreement with the National Science Foundation, on behalf of the Gemini Observatory partnership: the National Science Foundation (United States), National Research Council (Canada), Agencia Nacional de Investigaci\'{o}n y Desarrollo (Chile), Ministerio de Ciencia, Tecnolog\'{i}a e Innovaci\'{o}n (Argentina), Minist\'{e}rio da Ci\^{e}ncia, Tecnologia, Inova\c{c}\~{o}es e Comunica\c{c}\~{o}es (Brazil), and Korea Astronomy and Space Science Institute (Republic of Korea). 
\bibliography{references.bib}

\end{document}